\pdfoutput=1
\documentclass[aps,twocolumn,superscriptaddress]{revtex4-2}
\usepackage[colorlinks=true,citecolor=blue,urlcolor=blue,linkcolor=blue,pdfstartview=FitH,bookmarksopen]{hyperref}
\usepackage{amsmath,amssymb}
\usepackage{bm}
\usepackage{comment}
\usepackage{graphicx,xcolor}

\newcommand\+\dagger
\newcommand\p{{\bm{p}}}
\newcommand\q{{\bm{q}}}
\renewcommand\k{{\bm{k}}}

\renewcommand\v{\bm{v}}
\renewcommand\d\partial

\newcommand\<\langle
\renewcommand\>\rangle
\newcommand\M{\bm{\mathcal{M}}}
\newcommand\J{\bm{J}}
\renewcommand\Im{\mathop{\mathrm{Im}}}
\DeclareMathOperator{\arccosh}{arccosh}

\usepackage[normalem]{ulem}  

\newcommand{\diff}{\mathrm{d}}
\newcommand{\rme}{\mathrm{e}}
\newcommand{\rmi}{\mathrm{i}}

\begin{document}

\title{Universal Properties of Weakly Bound Two-Neutron Halo Nuclei}

\author{Masaru Hongo}
\affiliation{Department of Physics, University of Illinois, Chicago, Illinois 60607, USA}
\affiliation{RIKEN iTHEMS, RIKEN, Wako 351-0198, Japan}
\affiliation{Department of Physics, Niigata University, Niigata 950-2181, Japan}

\author{Dam Thanh Son}
\affiliation{Kadanoff Center for Theoretical Physics, University of Chicago, Chicago, Illinois 60637, USA}

\date{January 2022}

\begin{abstract}
  
We construct an effective field theory of a two-neutron halo nucleus
in the limit where the two-neutron separation energy $B$ and the
neutron-neutron two-body virtual energy $\epsilon_n$ are smaller than
any other energy scale in the problem, but the scattering between the
core and a single neutron is not fine-tuned, and the Efimov effect
does not operate.  The theory has one dimensionless coupling which
formally runs to a Landau pole in the ultraviolet.  We show that many
properties of the system are universal in the double fine-tuning
limit.  The ratio of the mean-square matter radius and charge radius
is found to be $\<r^2_m\>/\<r^2_c\>= A f(\epsilon_n/B)$, where $A$ is
the mass number of the core and $f$ is a function of the ratio
$\epsilon_n/B$ which we find explicitly.  In particular, when
$B\gg\epsilon_n$, $\< r^2_m\>/\<r^2_c\>= \frac23 A$.  The shape of
the $E1$ dipole strength function also depends only on the ratio
$\epsilon_n/B$ and is derived in explicit analytic form.  We estimate
that for the $^{22}$C nucleus higher-order corrections to our theory
are of the order of 20\% or less if the two-neutron separation energy is less
than 100~keV and the $s$-wave scattering length between a neutron and
a $^{20}$C nucleus is less than 2.8~fm.
\end{abstract}
\maketitle

\emph{Introduction.}---Neutron-rich nuclei near the neutron drip line
are at the forefront of modern nuclear physics.  Some of the most
exotic examples are two-neutron halo nuclei, consisting of a
relatively tightly bound core and two weakly bound neutrons, e.g.,
$^6$He, $^{11}$Li, and $^{22}$C.  These nuclei are called
``Borromean,'' i.e., bound states of three objects which would fall
apart when one is removed~\cite{Zhukov:1993aw}.

In this Letter, we develop an effective field theory (EFT) that can
describe Borromean two-neutron halo nuclei in the limit of very small
two-neutron separation energy $B$. The impetus to the construction of
this theory is the observation of the halo nucleus $^{22}$C with a
matter radius found to be as large as 5.4(9)~fm~\cite{Tanaka:2010}
which requires a small $B$: $B<100$~keV~\cite{Acharya:2013aea}.  A
later experiment~\cite{Togano:2016} yields a smaller matter radius---%
3.44(8)~fm---relaxing the upper limit to $B<400$~keV, but if one
incorporates the information about the neutron-core
scattering~\cite{Mosby:2013bix}, the upper limit is reduced to
$B<180$~keV~\cite{Hammer:2017tjm}.  So $^{22}$C is likely the least
bound among the known Borromean nuclei.

Our EFT requires two fine-tunings: We assume that the neutron-neutron
$s$-wave scattering length $a$ is unnaturally large and the
two-neutron separation energy of the halo nucleus is unnaturally
small.  In other words, the $n$-$n$ two-body virtual energy
$\epsilon_n=\hbar^2/(m_n a^2)\approx120$~keV (here, $m_n$ is the
neutron mass) and the binding energy $B$ of the core with two neutrons
are assumed to be smaller than all other energy scales in the problem.
We do not presume any hierarchy between these two energies.

Some previous attempts to apply the EFT philosophy to two-neutron halo
nuclei~\cite{Canham:2008jd,Hammer:2017tjm} rely on the existence of a
near-threshold resonance in the core-neutron subsystem and the
three-body Efimov effect~\cite{Naidon:2016dpf}.  This resonance seems
to be absent in the case of $^{22}$C, where experiment points to a
rather small $^{20}\text{C}+n$ scattering length~\cite{Mosby:2013bix}.
The theory developed in this Letter is designed to address this
situation.  It may also be a reasonable starting point for a
description of the $^3$He$^4$He$_2$ molecular trimer~\cite{Esry:1996},
whose binding energy 
($\sim 10$--$15$~mK)~\footnote{See, e.g.,
Ref.~\cite{Bressanini:2014}
for a compilation of theoretical
predictions.}
is somewhat smaller than the energy scale set by the
$^3\text{He}$--$^4\text{He}$ scattering length (around
50~mK~\cite{Uang:1982}).

The EFT, to be described, contains two relevant parameters and one
dimensionless coupling.  The two relevant parameters correspond to the
two fine-tunings.  The dimensionless coupling $g$ can be interpreted
as the probability that a halo nucleus splits into a core and a
two-neutron dimer (``dineutron''), and it runs logarithmically with
energy, reaching formally a Landau pole in the ultraviolet (UV) and
zero in the infrared (IR)~\footnote{In practice, the running is
limited in the UV by the UV cutoff and in the IR by $\epsilon_n$ or $B$,
whichever is larger.}.  All other coupling constants are irrelevant
and can be neglected in leading-order calculation.

Using the EFT, one can compute many physical quantities.  In particular,
we compute the ratio of the mean-square matter radius and charge
radius.  The result is particularly simple in the limit of infinite
neutron-neutron scattering length:
\begin{equation}
  \frac{\< r_m^2\>}{\<r_c^2\>} = \frac23 A,
\end{equation}
where $A$ is the mass number of the core.  We also obtain a fully
analytic expression for the $E1$ dipole strength function,
Eqs.~(\ref{E1}) and (\ref{fE1}).

All calculations in this Letter are performed under the assumption
that the core and the neutron are pointlike particles.  To translate
our results to the realistic nuclei, one needs account for the charge
and matter distribution inside the core and the neutrons.  In
addition, effects from irrelevant terms in the effective Lagrangian
may need to be taken into account.

\emph{The effective field theory.}---We first write down the effective
Lagrangian for the neutron sector.  Denoting the neutron by
$\psi_\alpha$, $\alpha=\uparrow,\downarrow$ being the spin index,
\begin{equation}\label{Ln}
  \mathcal L_n = \sum_\sigma  \psi^\+_\sigma 
   \left( \rmi \d_t + \frac{\nabla^2}{2m_n} \right) \psi_\sigma
  + c_0 \psi^\+_\uparrow \psi^\+_\downarrow \psi_\downarrow\psi_\uparrow.
\end{equation}
From now on, we set $m_n=1$.  Using a
Hubbard-Stratonovich transformation, the Lagrangian can be transformed
into
\begin{equation}
  \mathcal L_n = \sum_\sigma \psi^\+_\sigma 
   \left( \rmi \d_t + \frac{\nabla^2}2 \right) \psi_\sigma
  - \frac1{c_0} d^\+ d + \psi^\+_\uparrow \psi^\+_\downarrow d
  + d^\+ \psi_\downarrow \psi_\uparrow .
\end{equation}
\begin{figure}[t]
\begin{center}
\includegraphics[width=0.2\textwidth]{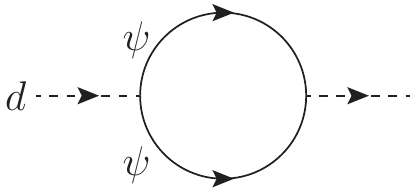}
  \caption{The self-energy of the dimer.}
\label{fig:dimer-prop}
\end{center}
\end{figure}
Computing the self-energy of the dimer $d$, which, in the
nonrelativistic theory, is exactly given by the one-loop diagram in
Fig.~\ref{fig:dimer-prop}, we find the full dimer propagator
\begin{equation}\label{dimerD}
  D(p) = - \frac{4\pi}{\sqrt{-p_0+\frac{\p^2}4}-\frac1a} \,,
\end{equation}
where $a$ denotes the $s$-wave scattering length given by
\begin{equation}
  \frac1{4\pi a} = - \frac1{c_0} + \int\!\frac{\diff \q}{(2\pi)^3}\,
  \frac1{\q^2} \,.
\end{equation}
The integral on the right-hand side linearly diverges in the UV and is
proportional to the UV cutoff.  The fine-tuning of $c_0$ leads to an
unnaturally large scattering length $a$.  Note that the UV behavior of
the dimer propagator is different from that of a free field.  In fact,
the UV behavior corresponds to a field of dimension 2:
$[d]=2$~\cite{Nishida:2007pj} \footnote{In our convention, the dimension of
momentum is 1 and energy is 2.}.

To construct the EFT describing the halo nucleus, we add into the
theory a field $\phi$ describing the core and $h$ describing the halo
nucleus.  They can be either bosonic or fermionic.  
The effective Lagrangian is now~\footnote{This Lagrangian was, in essence,
previously considered in Ref.~\cite{Son:2021kkx} for the case of a
three-body resonance, i.e., when the three-body binding
energy $B$ is negative.}
\begin{multline}\label{eff-L}
   \mathcal L = h^\+ \biggl( \rmi \d_t + \frac{\nabla^2}{2m_h} + B \biggr) h
   + \phi^\+ \biggl( \rmi \d_t + \frac{\nabla^2}{2m_\phi}\biggr)\phi\\
  + g (h^\+ \phi d + \phi^\+ d^\+ h) + \mathcal L_n
  + \text{counterterms},
\end{multline}
where $m_\phi=Am_n$ and $m_h=(A+2)m_n$ are the masses of the core and
the halo nucleus, respectively. As $[d]=2$ and
$[\phi]=[\psi]=\tfrac32$, the dimension of the interaction $h^\+ \phi
d$ is 5, which means that $g$ is dimensionless.  One can check that
terms not included in Eq.~(\ref{eff-L}) are all irrelevant, since they are accompanied by more fields or derivatives.  One can
compute the beta function for $g$~\cite{SM}:
\begin{equation}\label{beta-function}
  \frac{\d g}{\d \ln E} = \beta(g) =
  \frac2\pi \left( \frac A{A+2}\right)^{3/2} g^3 .
\end{equation}
The solution to this equation is
\begin{equation}\label{g2log}
  g^2(E) = \frac\pi4 \left( \frac{A+2}{A}\right)^{3/2} \frac1{\ln \frac{E_0}E}\,,
\end{equation}
where $E_0$ is the energy of the Landau pole.
Because of the properties of the nonrelativistic theory, our subsequent
calculations can be done to all orders in $g^2$.

One can arrive at the effective Lagrangian~(\ref{eff-L}) by starting
from a theory where the core $\phi$ and the resonantly interacting
neutron are coupled to each other by a contact interaction
$C_0 \phi^\+d^\+d\phi$, with a UV cutoff at the Landau pole scale.
Through a Hubbard-Stratonovich transformation, one introduces an
auxiliary field $h$ with the coupling $h^\+d\phi+\text{H.c}$.
Integrating out degrees of freedom in a energy shell between $E_0$ and
$E_1<E_0$, one generates a kinetic term for $h$ and arrive
to Eq.~(\ref{eff-L})~\footnote{In this scenario, the halo nucleus is bound
by the three-body force.  One should note, however, that the effective
Lagrangian is valid irrespective of the nature of the microscopic
force responsible for the binding of the halo nucleus.}.

\begin{figure}[b]
\begin{center}
\includegraphics[width=0.2\textwidth]{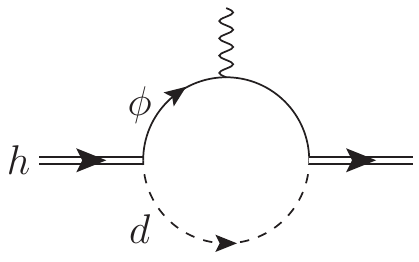}
  \caption{The Feynman diagram determining the charge form factor of
  the halo nucleus. The double line represents the halo nucleus, the
  single line---the core, and the dotted line---the neutron dimer, whose
  propagator is given in Eq.~(\ref{dimerD}).}
\label{fig:charge-ff}
\end{center}
\end{figure}

\emph{Charge and matter radii.}---We now proceed to extract physical
observables from the Lagrangian~(\ref{eff-L}).  The mean-square (rms)
charge radius (the rms of the deviation of the coordinates of the core
from the center of mass~\footnote{What we call here the mean-square
charge radius should be understood, for real nuclei, as the difference
between the mean-square point-proton radii of the halo and core
nuclei.  Similarly, what is later called the mean-square matter radius
is for real nuclei $\<r_m^2\>_h-\frac{A-2}A\<r_m^2\>_c$, where
$\<r_m^2\>_h$ and $\<r_m^2\>_c$ are the mean-square matter radii of
the halo and the core, respectively.}) 
can be extracted from the electric
form factor of the halo nucleus: $F(\k)= 1-\frac16 k^2 \<r_c^2\> + O(k^4)$ [recall that $F(\k)$ is a Fourier transform of the charge density].
The electric form factor is given by the Feynman diagram in
Fig.~\ref{fig:charge-ff}; it is proportional to
$g^2$, and by dimensional analysis one should have $ \< r_c^2 \> = g^2
B^{-1}f(\beta)$, where we introduce the dimensionless parameter
\begin{equation}
  \beta = \frac1{-a \sqrt{B}} = \sqrt{\frac{\epsilon_n}B} \,,
\end{equation}
where $\epsilon_n=1/a^2$ (we assume $a<0$). 
Computing the Feynman diagram~\cite{SM}, we find~\footnote{In subsequent
formulas, $g$ is the renormalized coupling in the on-shell
renormalization scheme.}
\begin{equation}\label{charge-radius}
  \< r_c^2 \>
  = \frac{ 4}\pi \frac{A^{1/2}}{(A+2)^{5/2}} \frac{g^2}B
  f_c( \beta ),
\end{equation}
where
\begin{equation}\label{fch2}
  f_c(\beta) = \begin{cases}
    \displaystyle{\frac1{1-\beta^2}} - \displaystyle{\frac{\beta\arccos\beta}{(1-\beta^2)^{3/2}}} \,, & \beta<1,\\
    & \\
    \displaystyle{-\frac1{\beta^2-1}} + \displaystyle{\frac{\beta\arccosh\beta}{(\beta^2-1)^{3/2}}}\,, &\beta>1.
    \end{cases}
\end{equation}

\begin{figure}[t]
\begin{center}
\includegraphics[width=0.2\textwidth]{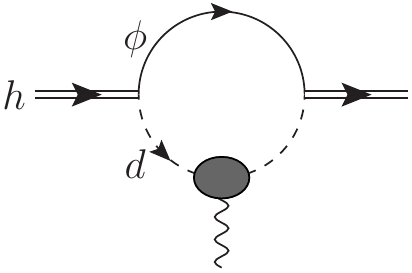}
  \caption{The Feynman diagram determining the ``neutron form factor''
  of the halo nucleus.}
\label{fig:neutron-ff}
\end{center}
\end{figure}

\begin{figure}[t]
\begin{center}
\includegraphics[width=0.48\textwidth]{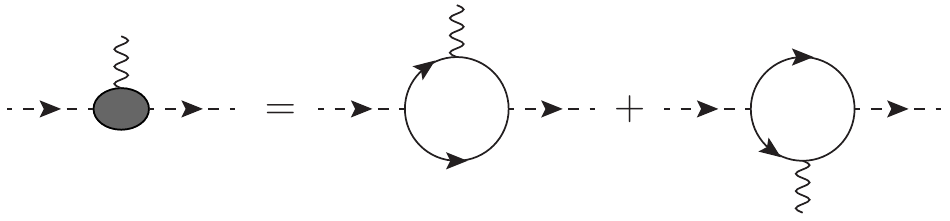}
  \caption{The effective vertex of the dimer-photon coupling.}
\label{fig:eff-vertex}
\end{center}
\end{figure}

One can further define the ``neutron radius'' by imagining that there
is a U(1) gauge boson coupled to the neutrons outside the
core~\footnote{This can be done by coupling a gauge field to $B-\frac
AZ Q$, where $B$ is the baryon charge, $Q$ is the electric charge, and
$Z$ and $A$ are the atomic and mass numbers, respectively, of the core.
}, which describes the spatial size of the dineutron distribution.  The
Feynman diagram determining the form factor of the halo nucleus with
respect to this ``neutron-number photon'' is drawn in
Fig.~\ref{fig:neutron-ff}, where the effective coupling of the dimer
to the photon is as in Fig.~\ref{fig:eff-vertex}.

The neutron radius is then
calculated to be~\cite{SM}
\begin{equation}
  \<r_n^2\> = \frac{g^2}{\pi B}
  \left(\frac{A}{A+2} \right)^{3/2}
  \biggl[ f_n(\beta) + \frac {A}{A+2} f_c(\beta) \biggr],
\end{equation}
where $f_c(\beta)$ is as in Eq.~(\ref{fch2})
and
\begin{equation}\label{fn}
  f_n(\beta) = 
  \begin{cases}
     \displaystyle{\frac1{\beta^3}} \biggl[ \pi-2\beta +
    (\beta^2-2)\displaystyle{
    \frac{\arccos\beta}{\sqrt{1-\beta^2}} }\biggr], & \beta <1,\\
       & \\
    \displaystyle{\frac1{\beta^3}} \biggl[  \pi-2\beta +
    (\beta^2-2)\displaystyle{
    \frac{\arccosh\beta}{\sqrt{\beta^2-1}} }\biggr], & \beta>1 .
    \end{cases}
\end{equation}
Both $f_c(\beta)$ and $f_n(\beta)$ are continuous at $\beta=1$ and
have the following asymptotics at small and large values of the
argument:
\begin{align}
  & f_c(0) = 1, \quad f_n(0) = \frac13\,,\\
  & f_c (\beta) 
  = \frac{\ln\beta}{\beta^2}, \quad 
  f_n (\beta)  = \frac{\ln\beta}{\beta^2},
  \quad \beta\to\infty.
\end{align}

From the charge radius and the neutron radius one can compute other
radii---the mean-square matter radius $\< r_m^2\>$, the neutron-neutron
distance $\<r_{nn}^2\>$, and the core-neutron distance $\<r_{cn}^2\>$
\footnote{
Using the positions of the core $\bm{r}_c$ and two neutrons $\bm{r}_1$ and $\bm{r}_2$, 
we define
$r_{m}^2 \equiv \frac{1}{A+2} (A \bm{r}_c^2 + \bm{r}_1^2 + \bm{r}_2^2)$, 
$r_{nn}^2 \equiv (\bm{r}_1 - \bm{r}_2)^2$, and 
$r_{cn}^2 \equiv \frac{1}{2} ((\bm{r}_1 - \bm{r}_c)^2 + (\bm{r}_2 - \bm{r}_c)^2)$.}:
\begin{align}
  \< r^2_m \> &= \frac{2}{A+2}  \< r_n^2 \> + \frac{A}{A+2} \< r_c^2 \>,\\
  \< r_{nn}^2\> &= 4\<r_n^2\> - A^2 \<r_c^2\>,\\
  \< r_{cn}^2\> &= 
 \<r_n^2\> + (A+1) \<r_c^2 \>.
\end{align}
When $\epsilon_n$ is fixed, these radii depend on $B$ in the following
way.  When $B\gg\epsilon_n$, the coupling $g$ is set at the scale $B$,
and $\<r^2\>\sim1/[B\ln(E_0/B)]$.  When $B\ll\epsilon_n$, $g$ is
frozen at the scale $\epsilon_n$, and the radii grow logarithmically
as $B\to0$: $\<r^2\>\sim\ln(\epsilon_n/B)$, which is a known
result~\cite{Fedorov:1994zz}.

Note that, due to the running of the coupling $g$, the results for the
radii are not truly ``universal'': They cannot be expressed solely in
terms of low-energy observables---the three-body binding energy $B$
and the neutron-neutron scattering length $a$.  Instead, they depend
logarithmically on the UV cutoff through the coupling $g$.  However,
the dependence on $g$ disappears when one computes the \emph{ratios}
of the radii.  For example, the ratio of the rms matter and charge
radii is
\begin{equation}\label{rmrc}
  \frac{\<r_m^2\>}{\<r_c^2\>}
  = \frac{A}2
  \biggl[ 1 + \frac{f_n(\beta)}{f_c(\beta)} \biggr],
\end{equation}
while the ratio of the core-neutron and neutron-neutron distances is
\begin{equation}\label{rcnrnn}
  \frac{\<r_{cn}^2\>}{\<r_{nn}^2\>} = \frac14 + \frac{A+2}{4A}
  \frac{f_c(\beta)}{f_n(\beta)}\,.
\end{equation}
In the two extreme limits $B\gg\epsilon_n$ and $B\ll\epsilon_n$,
these ratios become, respectively,
\begin{align}
  \frac{\<r_m^2\>}{\<r_c^2\>} &=
  \begin{cases}
     \tfrac23 A, & B \gg \epsilon_n,\\
     A, & B\ll \epsilon_n,
  \end{cases}\\
  \frac{\<r_{cn}^2\>}{\<r_{nn}^2\>} & =
  \begin{cases}
     1+\displaystyle{\frac3{2A_{\phantom{0}}}}\,, & B \gg \epsilon_n,\\
     \displaystyle{\frac{A+1^{\phantom{1}}}{2A}}\,, & B\ll \epsilon_n.
     \end{cases}
\end{align}
One notes, however, that $f_n/f_c$ reaches its large-$\beta$
limit very slowly.  For example, the ratio of the matter and charge
mean-square radii does not deviate more than 10\% from its
$B\gg\epsilon_n$ asymptotics unless $B$ is less than about
$\frac13\epsilon_n (\sim40~\text{keV})$.

\begin{figure}[t]
\begin{center}
\includegraphics[width=0.2\textwidth]{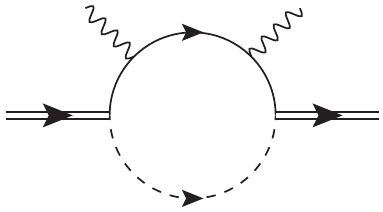}
  \caption{The Feynman diagram for the $E1$ dipole strength function.}
\label{fig:dis}
\end{center}
\end{figure}

\emph{The dipole strength function.}---The dipole strength function
can also be conveniently computed from EFT.  It is defined as
\begin{equation}
 \frac{\diff B(E1)}{\diff \omega} (\omega) =
  \sum_n |\<n| \bm{\mathcal M} |0\>|^2 \delta(E_n-E_0-\omega),
\end{equation}
where $|0\>$ is the ground state of the halo, the sum is taken over all excited
states $|n\>$, and $\bm{\mathcal M}$ is the dipole operator,
\begin{equation}
 \bm{\mathcal M} = \sqrt{\frac{3}{4\pi}} Ze
  (\bm{r}_c - \bm{R}_\text{c.m.}),
\end{equation}
where $\bm{r}_c$ is the coordinates of the core and
$\bm{R}_\text{c.m.}$ of the center of mass.  By noting that
\begin{equation}
 \frac\d{\d t} \M = \sqrt{\frac{3}{4\pi}} \J , 
\end{equation}
where $\J$ is the total electric current, one can rewrite the
dipole strength function as
\begin{equation}
 \frac{\diff B(E1)}{\diff \omega} = \frac{3}{4\pi}
  \frac1{\omega^2} \sum_n |\<n| \J |0\>|^2 \delta(E_n-E_0-\omega)
\end{equation}
and express it as the imaginary part of a two-point Green's function
of the current operator:
\begin{equation}
 \frac{\diff B(E1)}{\diff \omega} = -\frac{3}{4\pi}
  \frac1{\pi\omega^2} \Im G_{JJ}(\omega),
\end{equation}
where
\begin{equation}
 \rmi G_{JJ}(\omega)= 
  \!\int\!\diff t\, 
  \rme^{\rmi \omega t}
  \<0| T \J(t)\J(0) |0\> .
\end{equation}
The problem is now similar to that of deep inelastic scattering in
quantum chromodynamics~\cite{Peskin:1995ev}.  Computing the Feynman
diagram in Fig.~\ref{fig:dis}, we find~\cite{SM}
\begin{multline}\label{E1}
  \frac{\diff B(E1)}{\diff \omega} = \frac3{4\pi}Z^2e^2\frac{12g^2}{\pi}
  \frac{A^{1/2}}{(A+2)^{5/2}}
  \frac{(\omega-B)^2}{\omega^4}\\
  \times f_{E1}\biggl(\frac1{-a\sqrt{\omega-B}} \biggr),
\end{multline}
where
\begin{equation}\label{fE1}
  f_{E1}(x) = 1 - \frac83 x(1+x^2)^{3/2} + 4x^2 \left(1+\frac23 x^2 \right).
\end{equation}
The formula is more complicated than the formula for one-neutron halo
nuclei~\cite{Typel:2004zm} but is still explicit.

\begin{figure}[ht]
 \begin{center}
  \includegraphics[width=0.45\textwidth]{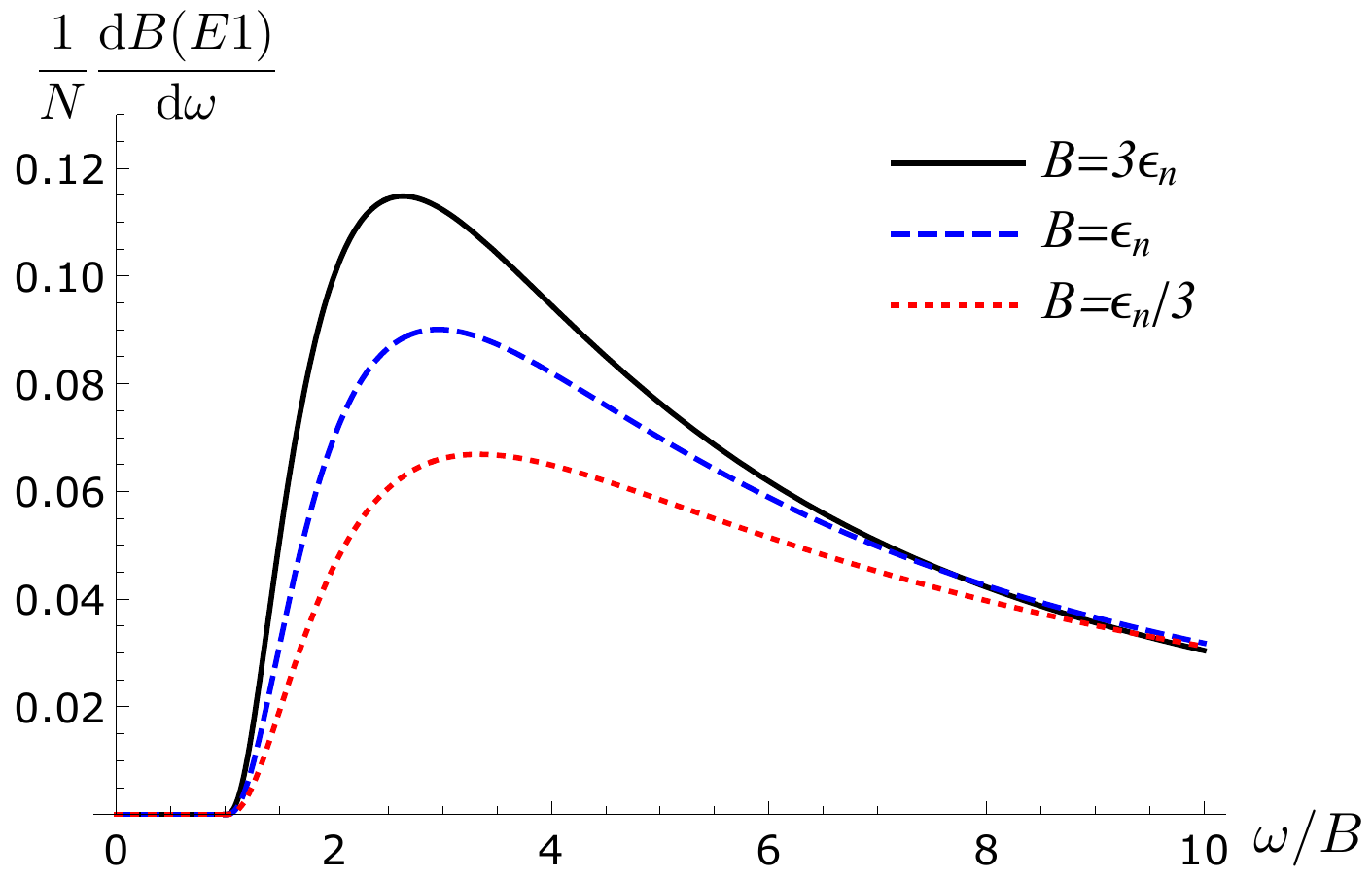}
  \caption{The $E1$ dipole strength function, plotted as function of
  $\omega/B$, for
  $B=3\epsilon_n$, $B=\epsilon_n$, and $B=\frac13\epsilon_n$.  
  The functions are so normalized by $N$ that the area under the theoretical
  curve, extended to $\omega/B=\infty$, is 1.}
  \label{fig:dipole}
 \end{center}
\end{figure}

One can check that the $E1$ dipole strength satisfies the sum rule
\begin{equation}
 \int\limits_0^\infty\! \diff \omega\, 
  \frac{\diff B(E1)}{\diff \omega}
  =  \frac3{4\pi} Z^2e^2 \< r_c^2\>,
\end{equation}
with the charge radius given by Eq.~(\ref{charge-radius}).  The
energy-weighted sum rule
\begin{equation}
  \int\limits_0^\infty\!\diff \omega\, \omega 
   \frac{\diff B(E1)}{\diff \omega}
  = \frac3{4\pi} Z^2 e^2 \frac3{A(A+2)} 
\end{equation}
is also valid if the logarithmic divergence of the integral on the
left-hand side is regularized by a UV cutoff at the energy of the
Landau pole.
The two sum rules are nontrivial checks of the self-consistency of our
theoretical approach.  The predicted shape of the $E1$ dipole strength
is plotted in Fig.~\ref{fig:dipole} as a function of $\omega/B$ for
various values of $\beta$.  One sees that the weight of the dipole
strength shifts to larger $\omega/B$ as $B/\epsilon_n$ decreases.

\emph{Applicability to real systems.}---The theory described above is
applicable when the binding energy of the halo $B$ and the $n$-$n$
two-body virtual energy $\epsilon_n$ are smaller than any other energy
scales in the problem.  In the real world, $\epsilon_n\approx0.12$~MeV
is indeed small.  For $^6$He and $^{11}$Li, the two-neutron separation
energy somewhat larger ($0.975$ and $0.369$~MeV, respectively); in
addition, the existence of near-threshold resonances in the $^5$He and
$^{10}$Li subsystem makes the applicability of our theory doubtful.

Nevertheless, let us try to compare our results with existing
experimental data and previous theoretical calculations.  For $^6$He,
Eq.~(\ref{rmrc}) predicts that $\<r_m^2\>/\<r_c^2\>\approx0.686A$.  In
Ref.~\cite{Danilin:2005fh}, it has been argued that the data for $^6$He
fit the formula $\<r_m^2\>/\<r_c^2\>=0.862A$, which the authors
derived approximately.  Our value is off by about 20\%.  For $^{11}$Li,
we compare our results with those of Ref.~\cite{Canham:2008jd}, where
$B=247$~keV and $\epsilon_n=116.04$~keV were used.  Setting the logarithm
in Eq.~(\ref{g2log}) to 1,
we find $\sqrt{\<r_c^2\>}=0.86~\text{fm}$ and
$\sqrt{\<r_n^2\>}=4.7~\text{fm}$, near the center of the error bands
predicted for large energies of the $^{10}$Li resonance.  The opening
angle $\theta_{nn}$ (defined as the vertex angle of the isosceles
triangle with sides $\sqrt{\<r_{cn}^2\>}$, $\sqrt{\<r_{cn}^2\>}$, and
$\sqrt{\<r_{nn}^2\>}$) is close to $60^\circ$ and is again within the
error band.  However, for the reasons listed above, it is possible
that the EFT provides only a qualitative guide for $^{11}$Li.

The theory presented here may be quantitatively useful for the
$^{22}$C nucleus if its two-neutron separation energy is indeed as
small as 100~keV~\cite{Acharya:2013aea}.  A correction to the EFT
comes from the scattering between the core and one neutron,
parametrized by the irrelevant dimension-6 term
$a_{cn}\phi^\+\psi^\+\psi\phi$.  The contributions from this term to
physical quantities should be suppressed by $a_{cn}(2m_n B)^{1/2}$
relative to the leading-order results, where $a_{cn}$ is the
core-neutron scattering length.  Experiment~\cite{Mosby:2013bix}
indicates that $|a_{cn}|<2.8$~fm, so this factor is $\le0.2$ (0.25 or
0.4 if the upper limit on $B$ is taken as 180 or 400~keV,
respectively).  Another dimension-6 operator, 
$d^\+(\rmi \d_t + \frac14\nabla^2)d$, 
has its coefficient fixed by the effective range
of the $s$-wave neutron-neutron scattering; its effect is expected to
be similarly suppressed.  Other terms, e.g., $h^\+ \nabla\phi\nabla
d$, have dimension 7 and higher and should be more suppressed.
Corrections from higher-order operators should be computable within
effective field theory.
The presence of the $^5$He $p$-wave resonance can be taken into account by adding the corresponding field~\cite{Bertulani:2002sz}. 
The present work is expected to open a potential direction to a quantitative study of the halo nuclei in addition to their universal properties.
For the $^3$He$^4$He$_2$ trimers we expect corrections from
$^3$He--$^4$He scattering to be relatively large: $(2\mu
B)^{1/2}|a_{34}|\sim 0.5$ (where $\mu$ is the reduced mass of the
$^3$He--$^4$He system and $a_{34}\approx -17$~\AA{} is the
$^3$He--$^4$He scattering length~\cite{Uang:1982}).  
Indeed, experiment and quantum Monte Carlo
simulations~\cite{Voigtsberger:2014,Bressanini:2014} seem to imply
substantially smaller values for the ratio $\<r_{cn}^2\>/\<r_{nn}^2\>$
compared the one given in Eq.~(\ref{rcnrnn}).

\acknowledgments

The authors thank Dario Bressanini, Reinhardt D\"orner, Maksim
Kunitski, and especially Hans-Werner Hammer for discussions. 
D.\,T.\,S.\ is supported, in part, by the U.S. Department of Energy Grant No.\ DE-FG02-13ER41958
and a Simons Investigator grant from the Simons Foundation.
M.\,H. is supported by the U.S. Department of Energy, Office of Science, Office of Nuclear Physics under Grant No.\ DE-FG0201ER41195 and partially by RIKEN iTHEMS Program (in particular, iTHEMS Non-Equilibrium Working Group and Mathematical Physics Working Group).

\bibliography{halo}

\begin{thebibliography}{29}%
\makeatletter
\providecommand \@ifxundefined [1]{%
 \@ifx{#1\undefined}
}%
\providecommand \@ifnum [1]{%
 \ifnum #1\expandafter \@firstoftwo
 \else \expandafter \@secondoftwo
 \fi
}%
\providecommand \@ifx [1]{%
 \ifx #1\expandafter \@firstoftwo
 \else \expandafter \@secondoftwo
 \fi
}%
\providecommand \natexlab [1]{#1}%
\providecommand \enquote  [1]{``#1''}%
\providecommand \bibnamefont  [1]{#1}%
\providecommand \bibfnamefont [1]{#1}%
\providecommand \citenamefont [1]{#1}%
\providecommand \href@noop [0]{\@secondoftwo}%
\providecommand \href [0]{\begingroup \@sanitize@url \@href}%
\providecommand \@href[1]{\@@startlink{#1}\@@href}%
\providecommand \@@href[1]{\endgroup#1\@@endlink}%
\providecommand \@sanitize@url [0]{\catcode `\\12\catcode `\$12\catcode
  `\&12\catcode `\#12\catcode `\^12\catcode `\_12\catcode `\%12\relax}%
\providecommand \@@startlink[1]{}%
\providecommand \@@endlink[0]{}%
\providecommand \url  [0]{\begingroup\@sanitize@url \@url }%
\providecommand \@url [1]{\endgroup\@href {#1}{\urlprefix }}%
\providecommand \urlprefix  [0]{URL }%
\providecommand \Eprint [0]{\href }%
\providecommand \doibase [0]{https://doi.org/}%
\providecommand \selectlanguage [0]{\@gobble}%
\providecommand \bibinfo  [0]{\@secondoftwo}%
\providecommand \bibfield  [0]{\@secondoftwo}%
\providecommand \translation [1]{[#1]}%
\providecommand \BibitemOpen [0]{}%
\providecommand \bibitemStop [0]{}%
\providecommand \bibitemNoStop [0]{.\EOS\space}%
\providecommand \EOS [0]{\spacefactor3000\relax}%
\providecommand \BibitemShut  [1]{\csname bibitem#1\endcsname}%
\let\auto@bib@innerbib\@empty
\bibitem [{\citenamefont {Zhukov}\ \emph {et~al.}(1993)\citenamefont {Zhukov},
  \citenamefont {Danilin}, \citenamefont {Fedorov}, \citenamefont {Bang},
  \citenamefont {Thompson},\ and\ \citenamefont {Vaagen}}]{Zhukov:1993aw}%
  \BibitemOpen
  \bibfield  {author} {\bibinfo {author} {\bibfnamefont {M.~V.}\ \bibnamefont
  {Zhukov}}, \bibinfo {author} {\bibfnamefont {B.~V.}\ \bibnamefont {Danilin}},
  \bibinfo {author} {\bibfnamefont {D.~V.}\ \bibnamefont {Fedorov}}, \bibinfo
  {author} {\bibfnamefont {J.~M.}\ \bibnamefont {Bang}}, \bibinfo {author}
  {\bibfnamefont {I.~J.}\ \bibnamefont {Thompson}},\ and\ \bibinfo {author}
  {\bibfnamefont {J.~S.}\ \bibnamefont {Vaagen}},\ }\bibfield  {title}
  {\bibinfo {title} {{Bound state properties of Borromean halo nuclei: $^6$He
  and $^{11}$Li}},\ }\href {https://doi.org/10.1016/0370-1573(93)90141-Y}
  {\bibfield  {journal} {\bibinfo  {journal} {Phys. Rep.}\ }\textbf {\bibinfo
  {volume} {231}},\ \bibinfo {pages} {151} (\bibinfo {year}
  {1993})}\BibitemShut {NoStop}%
\bibitem [{\citenamefont {Tanaka}\ \emph {et~al.}(2010)\citenamefont {Tanaka},
  \citenamefont {Yamaguchi}, \citenamefont {Suzuki}, \citenamefont {Ohtsubo},
  \citenamefont {Fukuda} \emph {et~al.}}]{Tanaka:2010}%
  \BibitemOpen
  \bibfield  {author} {\bibinfo {author} {\bibfnamefont {K.}~\bibnamefont
  {Tanaka}}, \bibinfo {author} {\bibfnamefont {T.}~\bibnamefont {Yamaguchi}},
  \bibinfo {author} {\bibfnamefont {T.}~\bibnamefont {Suzuki}}, \bibinfo
  {author} {\bibfnamefont {T.}~\bibnamefont {Ohtsubo}}, \bibinfo {author}
  {\bibfnamefont {M.}~\bibnamefont {Fukuda}}, \emph {et~al.},\ }\bibfield
  {title} {\bibinfo {title} {{Observation of a Large Reaction Cross Section in
  the Drip-Line Nucleus $^{22}\text{C}$}},\ }\href
  {https://doi.org/10.1103/PhysRevLett.104.062701} {\bibfield  {journal}
  {\bibinfo  {journal} {Phys. Rev. Lett.}\ }\textbf {\bibinfo {volume} {104}},\
  \bibinfo {pages} {062701} (\bibinfo {year} {2010})}\BibitemShut {NoStop}%
\bibitem [{\citenamefont {Acharya}\ \emph {et~al.}(2013)\citenamefont
  {Acharya}, \citenamefont {Ji},\ and\ \citenamefont
  {Phillips}}]{Acharya:2013aea}%
  \BibitemOpen
  \bibfield  {author} {\bibinfo {author} {\bibfnamefont {B.}~\bibnamefont
  {Acharya}}, \bibinfo {author} {\bibfnamefont {C.}~\bibnamefont {Ji}},\ and\
  \bibinfo {author} {\bibfnamefont {D.~R.}\ \bibnamefont {Phillips}},\
  }\bibfield  {title} {\bibinfo {title} {{Implications of a matter-radius
  measurement for the structure of Carbon-22}},\ }\href
  {https://doi.org/10.1016/j.physletb.2013.04.055} {\bibfield  {journal}
  {\bibinfo  {journal} {Phys. Lett. B}\ }\textbf {\bibinfo {volume} {723}},\
  \bibinfo {pages} {196} (\bibinfo {year} {2013})},\ \Eprint
  {https://arxiv.org/abs/1303.6720} {arXiv:1303.6720} \BibitemShut {NoStop}%
\bibitem [{\citenamefont {Togano}\ \emph {et~al.}(2016)\citenamefont {Togano}
  \emph {et~al.}}]{Togano:2016}%
  \BibitemOpen
  \bibfield  {author} {\bibinfo {author} {\bibfnamefont {Y.}~\bibnamefont
  {Togano}} \emph {et~al.},\ }\bibfield  {title} {\bibinfo {title}
  {{Interaction cross section study of the two-neutron halo nucleus
  $^{22}$C}},\ }\href {https://doi.org/10.1016/j.physletb.2016.08.062}
  {\bibfield  {journal} {\bibinfo  {journal} {Phys. Lett. B}\ }\textbf
  {\bibinfo {volume} {761}},\ \bibinfo {pages} {412} (\bibinfo {year}
  {2016})}\BibitemShut {NoStop}%
\bibitem [{\citenamefont {Mosby}\ \emph {et~al.}(2013)\citenamefont {Mosby}
  \emph {et~al.}}]{Mosby:2013bix}%
  \BibitemOpen
  \bibfield  {author} {\bibinfo {author} {\bibfnamefont {S.}~\bibnamefont
  {Mosby}} \emph {et~al.},\ }\bibfield  {title} {\bibinfo {title} {{Search for
  $^{21}$C and constraints on $^{22}$C}},\ }\href
  {https://doi.org/10.1016/j.nuclphysa.2013.04.004} {\bibfield  {journal}
  {\bibinfo  {journal} {Nucl. Phys.}\ }\textbf {\bibinfo {volume} {A909}},\
  \bibinfo {pages} {69} (\bibinfo {year} {2013})},\ \Eprint
  {https://arxiv.org/abs/1304.4507} {arXiv:1304.4507} \BibitemShut {NoStop}%
\bibitem [{\citenamefont {Hammer}\ \emph {et~al.}(2017)\citenamefont {Hammer},
  \citenamefont {Ji},\ and\ \citenamefont {Phillips}}]{Hammer:2017tjm}%
  \BibitemOpen
  \bibfield  {author} {\bibinfo {author} {\bibfnamefont {H.-W.}\ \bibnamefont
  {Hammer}}, \bibinfo {author} {\bibfnamefont {C.}~\bibnamefont {Ji}},\ and\
  \bibinfo {author} {\bibfnamefont {D.~R.}\ \bibnamefont {Phillips}},\
  }\bibfield  {title} {\bibinfo {title} {{Effective field theory description of
  halo nuclei}},\ }\href {https://doi.org/10.1088/1361-6471/aa83db} {\bibfield
  {journal} {\bibinfo  {journal} {J. Phys. G}\ }\textbf {\bibinfo {volume}
  {44}},\ \bibinfo {pages} {103002} (\bibinfo {year} {2017})},\ \Eprint
  {https://arxiv.org/abs/1702.08605} {arXiv:1702.08605} \BibitemShut {NoStop}%
\bibitem [{\citenamefont {Canham}\ and\ \citenamefont
  {Hammer}(2008)}]{Canham:2008jd}%
  \BibitemOpen
  \bibfield  {author} {\bibinfo {author} {\bibfnamefont {D.~L.}\ \bibnamefont
  {Canham}}\ and\ \bibinfo {author} {\bibfnamefont {H.-W.}\ \bibnamefont
  {Hammer}},\ }\bibfield  {title} {\bibinfo {title} {{Universal properties and
  structure of halo nuclei}},\ }\href
  {https://doi.org/10.1140/epja/i2008-10632-4} {\bibfield  {journal} {\bibinfo
  {journal} {Eur. Phys. J. A}\ }\textbf {\bibinfo {volume} {37}},\ \bibinfo
  {pages} {367} (\bibinfo {year} {2008})},\ \Eprint
  {https://arxiv.org/abs/0807.3258} {arXiv:0807.3258} \BibitemShut {NoStop}%
\bibitem [{\citenamefont {Naidon}\ and\ \citenamefont
  {Endo}(2017)}]{Naidon:2016dpf}%
  \BibitemOpen
  \bibfield  {author} {\bibinfo {author} {\bibfnamefont {P.}~\bibnamefont
  {Naidon}}\ and\ \bibinfo {author} {\bibfnamefont {S.}~\bibnamefont {Endo}},\
  }\bibfield  {title} {\bibinfo {title} {{Efimov physics: A review}},\ }\href
  {https://doi.org/10.1088/1361-6633/aa50e8} {\bibfield  {journal} {\bibinfo
  {journal} {Rep. Prog. Phys.}\ }\textbf {\bibinfo {volume} {80}},\ \bibinfo
  {pages} {056001} (\bibinfo {year} {2017})},\ \Eprint
  {https://arxiv.org/abs/1610.09805} {arXiv:1610.09805} \BibitemShut {NoStop}%
\bibitem [{\citenamefont {Esry}\ \emph {et~al.}(1996)\citenamefont {Esry},
  \citenamefont {Lin},\ and\ \citenamefont {Greene}}]{Esry:1996}%
  \BibitemOpen
  \bibfield  {author} {\bibinfo {author} {\bibfnamefont {B.~D.}\ \bibnamefont
  {Esry}}, \bibinfo {author} {\bibfnamefont {C.~D.}\ \bibnamefont {Lin}},\ and\
  \bibinfo {author} {\bibfnamefont {C.~H.}\ \bibnamefont {Greene}},\ }\bibfield
   {title} {\bibinfo {title} {Adiabatic hyperspherical study of the helium
  trimer},\ }\href {https://doi.org/10.1103/PhysRevA.54.394} {\bibfield
  {journal} {\bibinfo  {journal} {Phys. Rev. A}\ }\textbf {\bibinfo {volume}
  {54}},\ \bibinfo {pages} {394} (\bibinfo {year} {1996})}\BibitemShut
  {NoStop}%
\bibitem [{Note1()}]{Note1}%
  \BibitemOpen
  \bibinfo {note} {See, e.g., Ref.~\cite {Bressanini:2014} for a compilation of
  theoretical predictions.}\BibitemShut {Stop}%
\bibitem [{\citenamefont {Uang}\ and\ \citenamefont
  {Stwalley}(1982)}]{Uang:1982}%
  \BibitemOpen
  \bibfield  {author} {\bibinfo {author} {\bibfnamefont {Y.}~\bibnamefont
  {Uang}}\ and\ \bibinfo {author} {\bibfnamefont {W.~C.}\ \bibnamefont
  {Stwalley}},\ }\bibfield  {title} {\bibinfo {title} {{The possibility of a
  $^4$He$_2$ bound state, effective range theory, and very low energy He–He
  scattering}},\ }\href {https://doi.org/10.1063/1.442855} {\bibfield
  {journal} {\bibinfo  {journal} {J. Chem. Phys.}\ }\textbf {\bibinfo {volume}
  {76}},\ \bibinfo {pages} {5069} (\bibinfo {year} {1982})}\BibitemShut
  {NoStop}%
\bibitem [{Note2()}]{Note2}%
  \BibitemOpen
  \bibinfo {note} {In practice, the running is limited in the UV by the UV
  cutoff and in the IR by $\epsilon _n$ or $B$, whichever is
  larger.}\BibitemShut {Stop}%
\bibitem [{\citenamefont {Nishida}\ and\ \citenamefont
  {Son}(2007)}]{Nishida:2007pj}%
  \BibitemOpen
  \bibfield  {author} {\bibinfo {author} {\bibfnamefont {Y.}~\bibnamefont
  {Nishida}}\ and\ \bibinfo {author} {\bibfnamefont {D.~T.}\ \bibnamefont
  {Son}},\ }\bibfield  {title} {\bibinfo {title} {{Nonrelativistic conformal
  field theories}},\ }\href {https://doi.org/10.1103/PhysRevD.76.086004}
  {\bibfield  {journal} {\bibinfo  {journal} {Phys. Rev. D}\ }\textbf {\bibinfo
  {volume} {76}},\ \bibinfo {pages} {086004} (\bibinfo {year} {2007})},\
  \Eprint {https://arxiv.org/abs/0706.3746} {arXiv:0706.3746} \BibitemShut
  {NoStop}%
\bibitem [{Note3()}]{Note3}%
  \BibitemOpen
  \bibinfo {note} {In our convention, the dimension of momentum is 1 and energy
  is 2.}\BibitemShut {Stop}%
\bibitem [{Note4()}]{Note4}%
  \BibitemOpen
  \bibinfo {note} {This Lagrangian was, in essence, previously considered in
  Ref.~\cite {Son:2021kkx} for the case of a three-body resonance, i.e., when
  the three-body binding energy $B$ is negative.}\BibitemShut {Stop}%
\bibitem [{SM()}]{SM}%
  \BibitemOpen
  \href@noop {} {}\bibinfo {note} {{See Supplementary Material for
  details}}\BibitemShut {NoStop}%
\bibitem [{Note5()}]{Note5}%
  \BibitemOpen
  \bibinfo {note} {In this scenario, the halo nucleus is bound by the
  three-body force. One should note, however, that the effective Lagrangian is
  valid irrespective of the nature of the microscopic force responsible for the
  binding of the halo nucleus.}\BibitemShut {Stop}%
\bibitem [{Note6()}]{Note6}%
  \BibitemOpen
  \bibinfo {note} {What we call here the mean-square charge radius should be
  understood, for real nuclei, as the difference between the mean-square
  point-proton radii of the halo and core nuclei. Similarly, what is later
  called the mean-square matter radius is for real nuclei $\langle r_m^2\rangle
  _h-\protect \frac {A-2}A\langle r_m^2\rangle _c$, where $\langle r_m^2\rangle
  _h$ and $\langle r_m^2\rangle _c$ are the mean-square matter radii of the
  halo and the core, respectively.}\BibitemShut {Stop}%
\bibitem [{Note7()}]{Note7}%
  \BibitemOpen
  \bibinfo {note} {In subsequent formulas, $g$ is the renormalized coupling in
  the on-shell renormalization scheme.}\BibitemShut {Stop}%
\bibitem [{Note8()}]{Note8}%
  \BibitemOpen
  \bibinfo {note} {This can be done by coupling a gauge field to $B-\protect
  \frac AZ Q$, where $B$ is the baryon charge, $Q$ is the electric charge, and
  $Z$ and $A$ are the atomic and mass numbers, respectively, of the
  core.}\BibitemShut {Stop}%
\bibitem [{Note9()}]{Note9}%
  \BibitemOpen
  \bibinfo {note} {Using the positions of the core $\protect \bm {r}_c$ and two
  neutrons $\protect \bm {r}_1$ and $\protect \bm {r}_2$, we define $r_{m}^2
  \equiv \protect \frac {1}{A+2} (A \protect \bm {r}_c^2 + \protect \bm {r}_1^2
  + \protect \bm {r}_2^2)$, $r_{nn}^2 \equiv (\protect \bm {r}_1 - \protect \bm
  {r}_2)^2$, and $r_{cn}^2 \equiv \protect \frac {1}{2} ((\protect \bm {r}_1 -
  \protect \bm {r}_c)^2 + (\protect \bm {r}_2 - \protect \bm
  {r}_c)^2)$.}\BibitemShut {Stop}%
\bibitem [{\citenamefont {Fedorov}\ \emph {et~al.}(1994)\citenamefont
  {Fedorov}, \citenamefont {Jensen},\ and\ \citenamefont
  {Riisager}}]{Fedorov:1994zz}%
  \BibitemOpen
  \bibfield  {author} {\bibinfo {author} {\bibfnamefont {D.~V.}\ \bibnamefont
  {Fedorov}}, \bibinfo {author} {\bibfnamefont {A.~S.}\ \bibnamefont
  {Jensen}},\ and\ \bibinfo {author} {\bibfnamefont {K.}~\bibnamefont
  {Riisager}},\ }\bibfield  {title} {\bibinfo {title} {{Three-body halos: Gross
  properties}},\ }\href {https://doi.org/10.1103/PhysRevC.49.201} {\bibfield
  {journal} {\bibinfo  {journal} {Phys. Rev. C}\ }\textbf {\bibinfo {volume}
  {49}},\ \bibinfo {pages} {201} (\bibinfo {year} {1994})}\BibitemShut
  {NoStop}%
\bibitem [{\citenamefont {Peskin}\ and\ \citenamefont
  {Schroeder}(1995)}]{Peskin:1995ev}%
  \BibitemOpen
  \bibfield  {author} {\bibinfo {author} {\bibfnamefont {M.~E.}\ \bibnamefont
  {Peskin}}\ and\ \bibinfo {author} {\bibfnamefont {D.~V.}\ \bibnamefont
  {Schroeder}},\ }\href@noop {} {\emph {\bibinfo {title} {{An Introduction to
  Quantum Field Theory}}}}\ (\bibinfo  {publisher} {Addison-Wesley},\ \bibinfo
  {address} {Reading},\ \bibinfo {year} {1995})\BibitemShut {NoStop}%
\bibitem [{\citenamefont {Typel}\ and\ \citenamefont
  {Baur}(2004)}]{Typel:2004zm}%
  \BibitemOpen
  \bibfield  {author} {\bibinfo {author} {\bibfnamefont {S.}~\bibnamefont
  {Typel}}\ and\ \bibinfo {author} {\bibfnamefont {G.}~\bibnamefont {Baur}},\
  }\bibfield  {title} {\bibinfo {title} {{Effective-Range Approach and Scaling
  Laws for Electromagnetic Strength in Neutron-Halo Nuclei}},\ }\href
  {https://doi.org/10.1103/PhysRevLett.93.142502} {\bibfield  {journal}
  {\bibinfo  {journal} {Phys. Rev. Lett.}\ }\textbf {\bibinfo {volume} {93}},\
  \bibinfo {pages} {142502} (\bibinfo {year} {2004})},\ \Eprint
  {https://arxiv.org/abs/nucl-th/0406068} {arXiv:nucl-th/0406068} \BibitemShut
  {NoStop}%
\bibitem [{\citenamefont {Danilin}\ \emph {et~al.}(2005)\citenamefont
  {Danilin}, \citenamefont {Ershov},\ and\ \citenamefont
  {Vaagen}}]{Danilin:2005fh}%
  \BibitemOpen
  \bibfield  {author} {\bibinfo {author} {\bibfnamefont {B.~V.}\ \bibnamefont
  {Danilin}}, \bibinfo {author} {\bibfnamefont {S.~N.}\ \bibnamefont
  {Ershov}},\ and\ \bibinfo {author} {\bibfnamefont {J.~S.}\ \bibnamefont
  {Vaagen}},\ }\bibfield  {title} {\bibinfo {title} {{Charge and matter radii
  of Borromean halo nuclei: The $^6$He nucleus}},\ }\href
  {https://doi.org/10.1103/PhysRevC.71.057301} {\bibfield  {journal} {\bibinfo
  {journal} {Phys. Rev. C}\ }\textbf {\bibinfo {volume} {71}},\ \bibinfo
  {pages} {057301} (\bibinfo {year} {2005})}\BibitemShut {NoStop}%
\bibitem [{\citenamefont {Bertulani}\ \emph {et~al.}(2002)\citenamefont
  {Bertulani}, \citenamefont {Hammer},\ and\ \citenamefont
  {Van~Kolck}}]{Bertulani:2002sz}%
  \BibitemOpen
  \bibfield  {author} {\bibinfo {author} {\bibfnamefont {C.~A.}\ \bibnamefont
  {Bertulani}}, \bibinfo {author} {\bibfnamefont {H.~W.}\ \bibnamefont
  {Hammer}},\ and\ \bibinfo {author} {\bibfnamefont {U.}~\bibnamefont
  {Van~Kolck}},\ }\bibfield  {title} {\bibinfo {title} {{Effective field theory
  for halo nuclei: Shallow p-wave states}},\ }\href
  {https://doi.org/10.1016/S0375-9474(02)01270-8} {\bibfield  {journal}
  {\bibinfo  {journal} {Nucl. Phys.}\ }\textbf {\bibinfo {volume} {A712}},\
  \bibinfo {pages} {37} (\bibinfo {year} {2002})},\ \Eprint
  {https://arxiv.org/abs/nucl-th/0205063} {arXiv:nucl-th/0205063} \BibitemShut
  {NoStop}%
\bibitem [{\citenamefont {Voitgtsberger}\ \emph {et~al.}(2014)\citenamefont
  {Voitgtsberger} \emph {et~al.}}]{Voigtsberger:2014}%
  \BibitemOpen
  \bibfield  {author} {\bibinfo {author} {\bibfnamefont {J.}~\bibnamefont
  {Voitgtsberger}} \emph {et~al.},\ }\bibfield  {title} {\bibinfo {title}
  {{Imaging the structure of the trimer systems $^4$He$_3$ and
  $^3$He$^4$He$_2$}},\ }\href {https://doi.org/10.1038/ncomms6765} {\bibfield
  {journal} {\bibinfo  {journal} {Nat. Commun.}\ }\textbf {\bibinfo {volume}
  {5}},\ \bibinfo {pages} {5765} (\bibinfo {year} {2014})}\BibitemShut
  {NoStop}%
\bibitem [{\citenamefont {Bressanini}(2014)}]{Bressanini:2014}%
  \BibitemOpen
  \bibfield  {author} {\bibinfo {author} {\bibfnamefont {D.}~\bibnamefont
  {Bressanini}},\ }\bibfield  {title} {\bibinfo {title} {{The Structure of the
  Asymmetric Helium Trimer $^3$He$^4$He$_2$}},\ }\href
  {https://doi.org/10.1021/jp503090f} {\bibfield  {journal} {\bibinfo
  {journal} {J. Phys. Chem. A}\ }\textbf {\bibinfo {volume} {118}},\ \bibinfo
  {pages} {6521} (\bibinfo {year} {2014})}\BibitemShut {NoStop}%
\bibitem [{\citenamefont {Son}\ \emph {et~al.}(2021)\citenamefont {Son},
  \citenamefont {Stephanov},\ and\ \citenamefont {Yee}}]{Son:2021kkx}%
  \BibitemOpen
  \bibfield  {author} {\bibinfo {author} {\bibfnamefont {D.~T.}\ \bibnamefont
  {Son}}, \bibinfo {author} {\bibfnamefont {M.}~\bibnamefont {Stephanov}},\
  and\ \bibinfo {author} {\bibfnamefont {H.-U.}\ \bibnamefont {Yee}},\
  }\bibfield  {title} {\bibinfo {title} {{Fate of Multiparticle Resonances:
  From $Q$-Balls to $^3$He Droplets}},\ }\href@noop {} {\  (\bibinfo {year}
  {2021})},\ \Eprint {https://arxiv.org/abs/2112.03318} {arXiv:2112.03318}
  \BibitemShut {NoStop}%
\end{thebibliography}%

\clearpage
\onecolumngrid

\begin{center}
        \textbf{\large --- Supplementary Material ---\\ $~$ \\
        Universal Properties of Weakly Coupled Two-Neutron Halo Nuclei}\\
        \medskip
        \text{Masaru Hongo and Dam Thanh Son}
\end{center}
\setcounter{equation}{0}
\setcounter{figure}{0}
\setcounter{table}{0}
\setcounter{page}{1}
\makeatletter
\renewcommand{\thesection}{S\arabic{section}}
\renewcommand{\theequation}{S\arabic{equation}}
\renewcommand{\thefigure}{S\arabic{figure}}
\renewcommand{\bibnumfmt}[1]{[S#1]}

\section{Field theory: Feynman rules, renormalization}

In terms of the bare fields and bare couplings, the Lagrangian of the
theory of the halo nucleus is
\begin{equation}
  \mathcal L = h_0^\+ \left( \rmi \d_t + \frac{\nabla^2}{2m_h} + B_0 \right) h_0
  + \phi^\+ \left( \rmi \d_t + \frac{\nabla^2}{2m_\phi} \right) \phi
+ g_0 (h_0^\+ \phi d + \phi^\+ d^\+ h_0)
  + \mathcal L_n .
\end{equation}
where $\mathcal L_n$ is written in Eq.~(\ref{Ln}).  Define the
renormalized halo field $h$ and renormalized coupling $g$:
\begin{equation}
  h_0 = \sqrt{Z_h}\, h, \quad g = \sqrt{Z_h}\, g_0,
\end{equation}
the Lagrangian is
\begin{equation}
  \mathcal L = Z_h h^\+ \left( \rmi \d_t + \frac{\nabla^2}{2m_h} + B_0 \right) h
  + \phi^\+ \left( \rmi \d_t + \frac{\nabla^2}{2m_\phi} \right) \phi
  + g (h^\+ \phi d + \phi^\+ d^\+ h) + \mathcal L_n.
\end{equation}

The Feynman rules are as follows.  The dimer propagator is
\begin{equation}\label{Dfa}
 \rmi D(p) = - 4 \rmi \pi f_a \biggl( \frac{\p^2}4 - p_0 - \rmi \epsilon \biggr),
\end{equation}
where we introduce the notation
\begin{equation}\label{fa}
  f_a(x) = \frac1{\sqrt x- \frac1a} \,.
\end{equation}
The core propagator is
\begin{equation}
 \rmi G_\phi(p) = \frac{\rmi}{p_0 - \frac{\p^2}{2m_\phi}+ \rmi \epsilon} \,,
\end{equation}
and the halo-core-dimer vertex is $\rmi g$.

\begin{figure}[b]
\begin{center}
\includegraphics[width=0.27\textwidth]{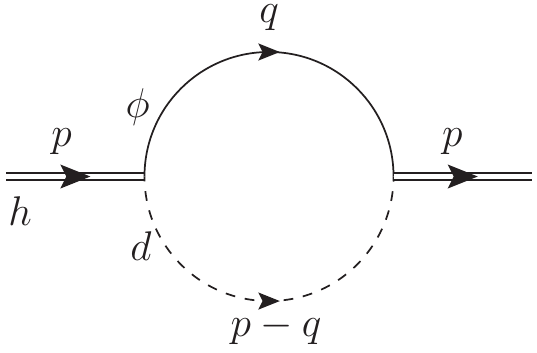}
  \caption{The self-energy of the halo nucleus.  The dash line in the
  loop is the dimer propagator, while the solid line is the propagator
  of the core.}
\label{fig:h-selfenergy}
\end{center}
\end{figure}

The self-energy of the $h$ field is given by a one-loop diagram,
(Fig.~\ref{fig:h-selfenergy})
\begin{equation}
  \Sigma(p) = \rmi (\rmi g)^2 \!\int\!\frac{\diff^4q}{(2\pi)^4}\,
   \rmi D(p-q) \rmi G_\phi(q)
   = -4 \pi \rmi g^2 \!\int\! \frac{\diff^4q}{(2\pi)^4}\,
  \frac{f_a(\frac14(\p-\q)^2 - p_0 + q_0 -\rmi \epsilon)}{q_0-\frac{\q^2}{2m_\phi}
  + \rmi \epsilon} \,.
\end{equation}
Closing the contour in the lower half-plane,
we find
\begin{equation}
  \Sigma(p) = - 4\pi g^2 \!\int\! \frac{\diff \q}{(2\pi)^3}\,
  f_a\biggl( -p_0 + \frac14(\p - \q)^2 + \frac{\q^2}{2m_\phi}
    \biggr).
\end{equation}
Performing a shift $\q\to\q+\frac{m_\phi}{m_h}\p$, we then get
\begin{equation}
  \Sigma(p) = - 4\pi g^2 \!\int\! \frac{\diff \q}{(2\pi)^3}\,
  f_a\biggl(-p_0 + \frac{\p^2}{2m_h} + \frac{\q^2}{2\mu} \biggr),
\end{equation}
where
\begin{equation}
  \mu = \frac{2m_\phi}{m_h} 
\end{equation}
is the reduced mass of the core-dimer system.

The full inverse propagator of the halo is then
\begin{equation}
  G_h^{-1}(p) = Z_h\biggl( p_0 - \frac{\p^2}{2m_h} + B_0 \biggr)
   + 4\pi g^2 \!\int\! \frac{\diff \q}{(2\pi)^3}\,
  f_a\biggl(-p_0 + \frac1{2m_h}\p^2 + \frac{\q^2}{2\mu} \biggr).
\end{equation}

The integral over $\q$ is quadratically divergent in the UV.  We
assume it is regularized by e.g., a momentum cutoff or by dimensional
regularization.  We use the on-shell renormalization scheme, where
the following two conditions are imposed:
\begin{align}
  G_h^{-1}(p_0,\mathbf{0})\bigl|_{p_0=-B} &= 0,\\
  \frac\d{\d p_0}G_h^{-1}(p_0, \mathbf{0})\Bigl|_{p_0=-B} &=1, 
\end{align}
which means
\begin{align}
  Z_h(B_0-B) +  4\pi g^2\!\int\! \frac{\diff \q}{(2\pi)^3}\,
  f_a (B_\q) &= 0,\\
  Z_h - 4\pi g^2 \!\int\! \frac{\diff \q}{(2\pi)^3}\,
  f_a' ( B_\q ) &= 1, \label{onshell-ren2}
\end{align}
where
\begin{equation}
  B_\q =  B + \frac{\q^2}{2\mu} \,,
\end{equation}
and the function $f_a$ is defined as in Eq.~(\ref{fa}).

From Eq.~(\ref{onshell-ren2}) one can derive the beta function for the
running of the coupling $g$.  Noting that $Z=g^2/g_0^2$, we find
\begin{equation}
  g_0^2 = \frac{g^2}{1 + 4\pi g^2 \! \int \! \frac{\diff \q}{(2\pi)^3}\,
  f_a' ( B_\q )} \,.
\end{equation}
Differentiating $g_0$ with respect to the the UV cutoff in the integral,
we find the beta function~(\ref{beta-function}).

\section{Charge radius}

\begin{figure}[t]
\begin{center}
\includegraphics[width=0.4\textwidth]{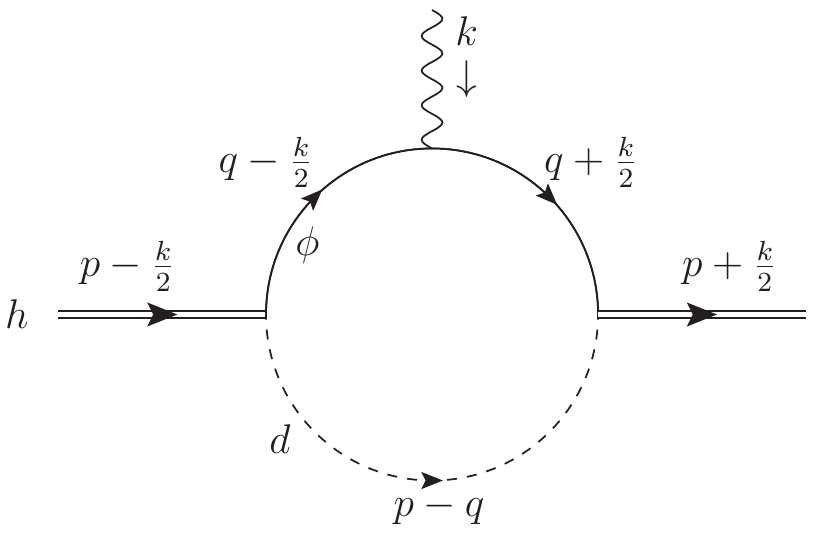}
  \caption{The Feynman diagram that determines the charge form factor of
  the halo nucleus.}
\label{fig:charge-ff-pqk}
\end{center}
\end{figure}

The charge form factor of the halo nucleus is
\begin{equation}\label{FZGamma}
  F(\k) = Z_h + \Gamma(k,p) ,
\end{equation}
where $\Gamma (k,p)$ is given by the Feynman diagram in Fig.~\ref{fig:charge-ff-pqk} 
at the kinematic point
\begin{equation}\label{kinematic-point}
  k = (0, \k),\quad
  p = \biggl( -B + \frac{\k^2}{8m_h}, \bm{0} \biggr).
\end{equation}

We now compute the Feynman diagram in Fig.~\ref{fig:charge-ff-pqk}.
Using the Feynman rules we find
\begin{align}
 \Gamma (k,p)
 &= (\rmi g)^2 \!\int\! \frac{\diff^4 q}{(2\pi)^4}\,
 \rmi G_{\phi} \left( q - \frac k2 \right) 
 \rmi G_{\phi} \left( q+ \frac k2\right)
 \rmi D(p-q)
 \nonumber \\
 &= \rmi g^2\!\int\!\frac{\diff^4 q}{(2\pi)^4}\,
 \frac{D\left(-B +\frac{\k^2}{8m_h} - q_0, -\q\right)}
 {\left(q_0 -\varepsilon_{\q-\frac\k2} + \rmi \epsilon\right)
 \left(q_0  -\varepsilon_{\q+\frac\k2} + \rmi \epsilon\right)} \,,
\end{align}
where $\varepsilon_\q=\q^2/(2m_\phi)$.  Performing integral by
closing the contour in the lower half-plane, one finds
\begin{equation}\label{Gamma-unexp}
 \Gamma (k,p) = g^2\!\int\!\frac{\diff \q}{(2\pi)^3} \,
  \frac1{\varepsilon_{\q-\frac\k2}-\varepsilon_{\q+\frac\k2}}
  \left[ D \left(-B+\frac{\k^2}{8m_h}-\varepsilon_{\q-\frac\k2} , -\q\right)
   - D \left(-B+\frac{\k^2}{8m_h}-\varepsilon_{\q+\frac\k2} , -\q\right)
  \right].
\end{equation}
Using Eq.~(\ref{Dfa}), we have
\begin{equation}
  F(\k) = Z_h + 
  4\pi g^2\!\int\!\frac{\diff \q}{(2\pi)^3} \,
  \frac{m_\phi}{\q\cdot\k}
  \left[f_a \left(B_\q +\frac{\k^2}{4m_\phi m_h}
    - \frac{\q\cdot\k}{2m_\phi}\right)
    -f_a \left(B_\q +\frac{\k^2}{4m_\phi m_h}
    +\frac{\q\cdot\k}{2m_\phi} \right) \right].
\end{equation}
To order $k^0$  we have
\begin{equation}
  F(0) = Z_h - 4\pi g^2 \!\int\!\frac{\diff \q}{(2\pi)^3} \,
  f_a' ( B_\q ) = 1 .
\end{equation}
where we have used Eq.~(\ref{onshell-ren2}).  This is just the
statement that the total charge is equal to 1
in our normalization.

Expanding Eq.~(\ref{Gamma-unexp}) to second order in $\k$, we
then find the charge radius from 
$F(\k)=1- \frac{1}{6} \<r_c^2\> k^2 + O(k^4)$:
\begin{equation}
  \< r_c^2 \> =
  \frac{4\pi g^2}{m_\phi m_h} \!
  \int\!\frac{\diff \q}{(2\pi)^3} \left[\frac32 f_a'' (B_\q)
    + \frac{\q^2}{6\mu} 
    f_a''' (B_\q ) \right],
\end{equation}
which can be written as
\begin{equation}
  \< r_c^2 \>
  = \frac{ 4g^2}\pi \frac{A^{1/2}}{(A+2)^{5/2}} \frac{1}B
  f_c \left( \frac1{(-a)\sqrt B} \right),
\end{equation}
where
\begin{equation}
  f_c(\beta) =  \int\limits_1^\infty\!\diff y \, \sqrt{y-1}
  \left[ 3 f_a''(y) + \frac23 (y-1) f_a'''(y)\right]\biggl|_{a=-1/\beta}
\end{equation}
Integrating by part, this can be reduced into the form
\begin{equation}
  f_c(\beta) = -\! \int\limits_1^\infty\!\diff y \,
      \frac{f_a'(y)|_{a=-1/\beta}}{\sqrt{y-1}}
  = \int\limits_1^\infty\!\diff y \,
  \frac1{2\sqrt{y(y-1)} (\sqrt{y}+\beta)^2}
\end{equation}
Evaluating the integral, one obtains
Eq.~(\ref{fch2}).  The result for the charge radius also coincides
with the value obtained from a sum rule for the dipole strength
function, computed in Section~\ref{SM:E1}.

\section{Neutron radius}

\subsection{Effective coupling of the dimer to the ``neutron-number photon''}
\label{sec:Gamma_ddg}

\begin{figure}[htb]
\begin{center}
 \includegraphics[width=0.85\textwidth]{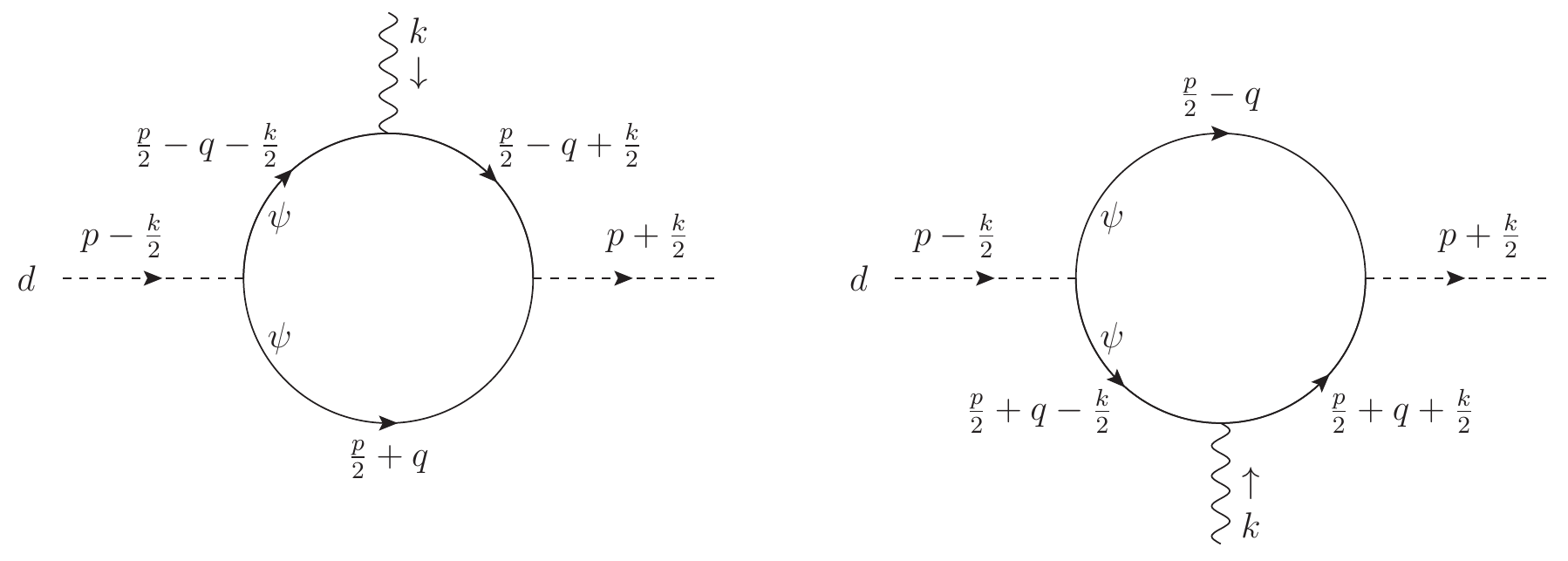}
  \caption{The diagrams contributing to the effective coupling of a ``neutron-number photon'' to the dimer field.}
\label{fig:effective-vertex}
\end{center}
\end{figure}

The two diagrams contributing to the effective coupling of the dimer
to the ``neutron-number photon'' are depicted in
Fig.~\ref{fig:effective-vertex}.  They sum up to
\begin{equation}
  \Gamma_{dd\gamma} 
   (k,p) 
   = \rmi^2 \!\int\! \frac{\diff^4 q}{(2\pi)^4}\,
  \rmi G\left(\frac p2 - q - \frac k2 \right)
  \rmi G\left(\frac p2 - q + \frac k2 \right)
  \rmi G\left(\frac p2 + q  \right) + (q\to -q).
\end{equation}
Performing the integral over $q_0$, this becomes
\begin{equation}
  \Gamma_{dd\gamma} (k,p)
   = \int\!\frac{\diff \q}{(2\pi)^3}\,
  \frac2{ \left( p_0 -\frac{\p^2}4 -\q^2 - \frac{\k^2}8 \right)^2
  -\frac14\left(k_0 - \frac{\p\cdot\k}2 + \q\cdot\k \right)^2} \,.
\end{equation}
Expanding to quadratic order in $\k$ and $k_0$, 
one can write the result in terms of  the three Galilean invariant quantities
\begin{equation}
  P_0 = p_0 - \frac{\p^2}4 \,, \qquad K_0 = k_0 - \frac{\p\cdot\k}2\,,
  \qquad
  k = |\k|,
\end{equation}
as
\begin{equation}
  \Gamma_{dd\gamma} (k,p)
   = \Gamma_0(P_0) + k^2 \Gamma_1(P_0) + K_0^2 \Gamma_2(P_0),
\end{equation}
where
\begin{align}
  \Gamma_0(P_0) & = \int\!\frac{\diff \q}{(2\pi)^3}\,
  \frac2{\left( P_0 -\q^2 \right)^2}
  = \frac1{4\pi}\frac{1}{\sqrt{-P_0}} \,,\\
  \Gamma_1(P_0) &= \int\! \frac{\diff \q}{(2\pi)^3}
  \left[ \frac12 \frac1{(P_0-\q^2)^3} + \frac16 \frac{\q^2}{(P_0-\q^2)^4}\right]   = -\frac{5}{384\pi}\frac1{(-P_0)^{3/2}} \,, \\
  \Gamma_2(P_0) &= \frac12\int\!\frac{\diff \q}{(2\pi)^3}\,
     \frac1{\left( P_0 -\q^2 \right)^4}
  =\frac1{128\pi} \frac1{(-P_0)^{5/2}} \,.   
\end{align}

\subsection{Calculation of the neutron radius}

\begin{figure}[t]
\begin{center}
\includegraphics[width=0.4\textwidth]{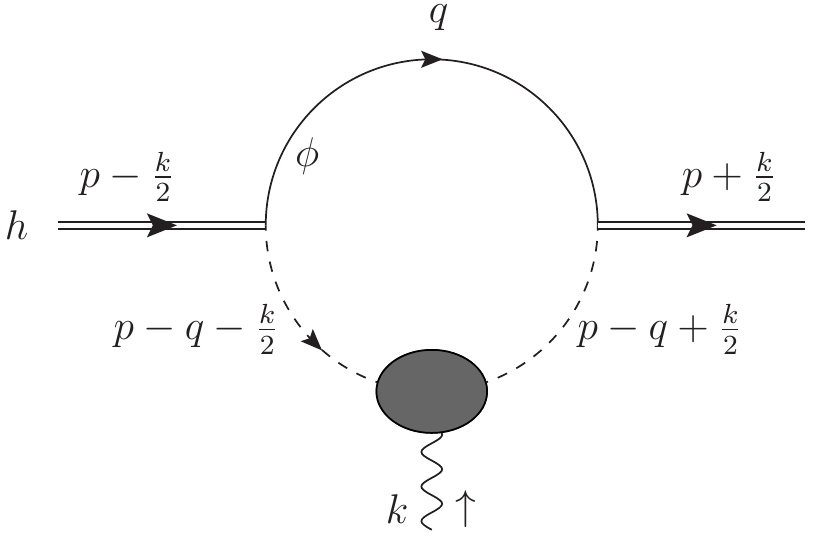}
  \caption{The Feynman diagram that contributes to the neutron
  form factor of the halo nucleus.}
\label{fig:n-ff-pqk}
\end{center}
\end{figure}

We imagine that there exists a gauge boson that couples to the
neutrons outside the core, but not to the core.  The ``neutron form
factor'' receives a contribution from the Feynman diagram in
Fig.~\ref{fig:n-ff-pqk}, and defined by 
\begin{equation}
 F_n (\k) =  2Z_h + \Gamma_n (k,p),
\end{equation}
where $\Gamma_n (k,p)$ is given by 
\begin{equation}
 \Gamma_n (k,p) = \rmi g^2 \!\int\!\frac{\diff^4 q}{(2\pi)^4}\,
  G_\phi(q) D\left(p-q-\frac k2\right)\Gamma_{dd\gamma}(k,p-q)
  D\left(p-q+\frac k2\right)
\end{equation}
at the kinematic point (\ref{kinematic-point}).
Here $\Gamma_{dd\gamma}$ is the effective vertex of the coupling of
the dimer to the ``neutron-number photon.''  This vertex has been
evaluated in Sec.~\ref{sec:Gamma_ddg} to second order in the photon
momentum to
\begin{equation}
  \Gamma_{dd\gamma}(k,p) = \frac1{4\pi}\left[
  \frac{1}{\sqrt{-P_0}}
  -\frac 5{96} \frac{ \k^2}{(-P_0)^{3/2}}
  + \frac 1{32} \frac{K_0^2}{(-P_0)^{5/2}} \right],
\end{equation}
where $P_0=p_0-\p^2/4$, $K_0=k_0-\frac12\p\cdot\k$.

Integrating over the $q_0$, closing the contour in the lower
half-plane, one picks up the pole from the core propagator $G_\phi(q)$
\begin{equation}
 \Gamma_n (k,p)
 = g^2 \!\int\!\frac{\diff \q}{(2\pi)^3}\,
  D\left(p-q-\frac k2\right)\Gamma_{dd\gamma}(k,p-q)
  D\left(p-q+\frac k2\right) \biggl|_{q_0=\frac1{2m_\phi}\q^2} \,.
\end{equation}  
At the kinematic point~(\ref{kinematic-point}), the neutron form
factor of the halo nucleus is, to quadratic order in $\k$,
\begin{align}
 F_n (\k)
 = 2Z_h + 4\pi g^2 \!\int\!\frac{\diff \q}{(2\pi)^3}
 &\left[
 \frac{1}{\sqrt{B_\q}}
 + \left(-\frac5{96} + \frac{1}{16m_h} \right)
 \frac{\k^2}{B_\q^{3/2}}
 + \frac{1}{128}\frac{(\k\cdot\q)^2}{B_\q^{5/2}}
 \right]
 \nonumber \\
 \times 
 &\left[
 \frac1{\left( \sqrt{B_\q} - \frac1{a} \right)^2}
 -\frac{m_\phi}{16m_h} \frac{\k^2}{\left( \sqrt{B_\q} - \frac1{a}\right)^3
 \sqrt{B_\q}}
 +\frac1{64} \frac{\left( 2\sqrt{B_\q} - \frac1{a}\right) (\q\cdot\k)^2}
 { \left( \sqrt{B_\q} - \frac1{a}\right)^4 B_\q^{3/2} }
 \right] .
\end{align}

Recalling Eq.~\eqref{onshell-ren2}, we find that $F_n (0)$ becomes
\begin{equation}
 F_n (0) 
  = 2 \left[
   Z_h - 4 \pi g^2\!\int\!\frac{\diff\q}{(2\pi)^3}\, f'(B_\q) 
  \right]
  = 2 ,
\end{equation}
which is the total ``neutron number'' of the halo nucleus.
Computing $F_n (\k)$ to order $\k^2$, we find the neutron radius,
\begin{equation}
  \<r_n^2\> = \frac{g^2}{\pi B}
  \left(\frac{A}{A+2} \right)^{3/2}
  \left[ f_n(\beta) + \frac A{A+2} f_c(\beta) \right],
\end{equation}
where $f_c(\beta)$ is as in Eq.~(\ref{fch2}),
and
\begin{equation}
  f_n(\beta) = \int\!dy\, \frac{\sqrt{y-1}}{2y^{3/2}(\sqrt y+\beta)^2}
\end{equation}
Taking the integral, one obtains Eq.~(\ref{fn}).

\section{$E1$ dipole strength function}
\label{SM:E1}

\begin{figure}[htb]
\begin{center}
\includegraphics[width=0.36\textwidth]{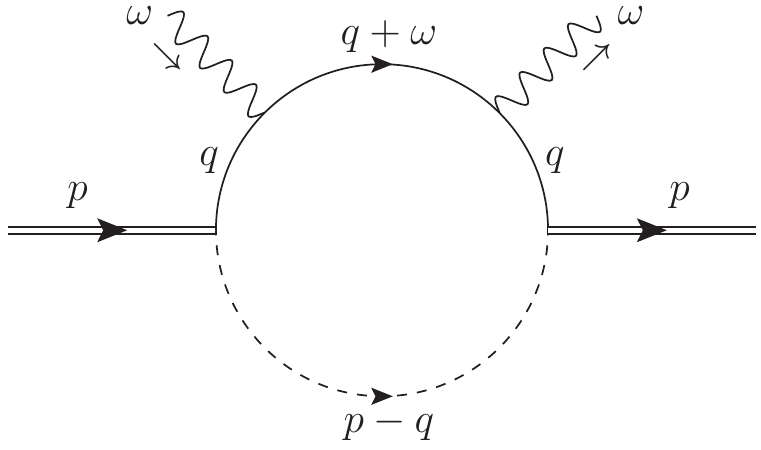}
  \caption{The Feynman diagram determining the $E1$ dipole strength
  function.  A second diagram obtained by reversing the direction of
  momentum flow on the two photon lines contributes to $G_{JJ}$ but
  not to its imaginary part when $\omega>0$.}
\label{fig:dis-momentum}
\end{center}
\end{figure}

To find the $E1$ dipole strength function, one needs to evaluate the
Feynman diagrams, one of which is drawn on
Fig.~\ref{fig:dis-momentum}
\begin{equation}
 \rmi G_{JJ} (\omega) =
  (Ze)^2
  (\rmi g)^2\! \int \! \frac{\diff^4 q}{(2\pi)^4}\,
  \rmi G_{\phi} (q) \frac{\q}{m_\phi} 
  \rmi G_{\phi} (q+\omega) \frac{\q}{m_\phi} 
  \rmi G_{\phi} (q) \rmi D(p-q) + (\omega\to-\omega),
\end{equation}
where $\omega=(\omega,\mathbf{0})$.  Closing the contour in the lower
half-plane, the imaginary part comes from the pole in $G(q+\omega)$.
\begin{align}
 \Im G_{JJ}(\omega) 
 &= (Ze)^2
  \frac{g^2}{m_\phi^2\omega^2} \!\int\! \frac{\diff \q}{(2\pi)^3}\,
  \q^2 \Im D\left(\omega-B-\frac{\q^2}{2m_\phi}, -\q\right) 
 \nonumber \\
 &= - (Ze)^2
 \frac{4\pi g^2}{m_\phi^2\omega^2}\!\int\! \frac{\diff \q}{(2\pi)^3}\,
 \q^2 \frac{\sqrt{\omega-B-\frac{\q^2}{2\mu}}}
 {\omega-B-\frac{\q^2}{2\mu}+\frac1{a^2}}
 \theta\left(\omega-B-\frac{\q^2}{2\mu}\right).
\end{align}
Evaluating the integral one finds
\begin{equation}
  \Im G_{JJ}(\omega) = - (Ze)^2
  \frac{3g^2}{8}\frac{(2\mu)^{5/2}}{m_\phi^2}
  \frac{(\omega-B)^2}{\omega^2}
  f_{E1}\left(\frac1{(-a)\sqrt{\omega-B}} \right),
\end{equation}
with the function $f_{E1}(x)$ defined in Eq.~(\ref{fE1}). 
From this one obtains Eq.~(\ref{E1}).

\section{Relationships between various mean square radii}

Let $\bm{r}_c$ denote the position of the core, and $\bm{r}_1$ and
$\bm{r}_2$ those of the two neutrons.  
Assume that the center of mass is at the origin,
\begin{equation}
  A\bm{r}_c + \bm{r}_1 +\bm{r}_2 = \bm{0},
   \label{eq:center-of-mass}
\end{equation}
then the coordinates of every particle can be express through
$\bm{r}_c$ and $\bm{r}_{nn}= \bm{r}_1-\bm{r}_2$:
\begin{align}
  \bm{r}_1 &= -\frac A2 \bm{r}_c + \frac12 \bm{r}_{nn} ,\\
  \bm{r}_2 &= -\frac A2 \bm{r}_c - \frac12 \bm{r}_{nn} .
\end{align}
Now notice that $\< \bm{r}_c\cdot \bm{r}_{nn}\> = 0$ due to the symmetry of the ground-state wavefunction of the halo with respect to exchanging $\bm{r}_1$ and $\bm{r}_2$, one can derive relationships between different
mean-square radii.
For example
\begin{equation}
  \< r_n^2\> = \< \bm{r}_1^2\> = \frac{A^2}4\<r_c^2\>
  + \frac14\<r_{nn}^2\> 
  ~\Rightarrow~ \<r_{nn}^2\> = 4\<r_n^2\>-A^2\<r_c^2\>.
\end{equation}
Analogously
\begin{align}
 \< r_m^2\> 
 &= \frac1{A+2} \big( A\<r_c^2\> + \< \bm{r}_1^2\> + \< \bm{r}_2^2\> \big)
 = \frac{2}{A+2} \< r_n^2\> + \frac{A}{A+2} \<r_c^2\>,
 \\
 \< r_{cn}^2\> 
 &= 
 \frac{1}{2} 
 \left[ \<(\bm{r}_1 - \bm{r}_c)^2 \> + \<(\bm{r}_2 - \bm{r}_c)^2 \> \right]
 = \<r_n^2\> + (A+1) \<r_c^2\> ,
\end{align}
where we used Eq.~\eqref{eq:center-of-mass} to derive the second relation.

\end{document}